\documentclass[useAMS,usenatbib]{mn2e} 
\usepackage{graphicx} 
\title{Constraining dark matter by the 511 keV line} 
\author[Chan and Leung]{Man Ho Chan \thanks{chanmh@eduhk.hk}, Chung Hei Leung
\\ Department of Science and Environmental Studies, The Education University of Hong Kong, Tai Po, Hong Kong}

\begin{document}

\date{Accepted XXXX, Received XXXX}

\pagerange{\pageref{firstpage}--\pageref{lastpage}} \pubyear{XXXX}

\maketitle

\label{firstpage}

\date{\today}

\begin{abstract}
In the past few decades, observations indicated that an unexplained high production rate of positrons (the strong 511 keV line) exists in the Milky Way center. By using the fact that a large amount of high density gas used to exist near the Milky Way center million years ago, we model the rate of positrons produced due to dark matter annihilation. We consider the effect of adiabatic contraction of dark matter density due to the supermassive black hole at the Milky Way center and perform a detailed calculation to constrain the possible annihilation channel and dark matter mass range. We find that only three annihilation channels ($\mu^+\mu^-$, $4e$ and $4 \mu$) can provide the required positron production rate and satisfy the stringent constraint of gamma-ray observations. In particular, the constrained mass range for the $\mu^+\mu^-$ channel is $m \approx 80-100$ GeV, which is close to the mass range obtained for the dark matter interpretation of the GeV gamma-ray and positron excess. In other words, the proposed scenario can simultaneously provide the required positron production rate to explain the 511 keV emission, the positron excess and the GeV gamma-ray excess in our Milky Way, and it is compatible with the density spike due to adiabatic growth model of the supermassive black hole.
\end{abstract}


\section{Introduction}
Observations of soft gamma-ray indicate a strong flux of 511 keV photons $\phi_{511} \sim 10^{-3}$ ph cm$^{-2}$ s$^{-1}$ emitted in the Milky Way \citep{Leventhal,Knodlseder}. These 511 keV photons originate from the positrons produced in the bulge and disk with production rate $\dot{N}_{e^+}=11.5^{+1.8}_{-1.44} \times 10^{42}$ s$^{-1}$ and $\dot{N}_{e^+}=8.1^{+1.5}_{-1.4} \times 10^{42}$ s$^{-1}$ respectively \citep{Prantzos,Perets}. This abnormally high bulge to disk ratio $B/D=1.42^{+0.34}_{-0.30}$ is difficult to explain \citep{Prantzos,Perets}. Many different astrophysical processes have been suggested to account for the 511 keV line (see the review in \citet{Prantzos}). However, none of them is successful. The models of supernovae, x-ray binaries or microquasars can explain only about half of the strong 511 keV emission from the inner Milky Way \citep{Prantzos}. Therefore, including the contributions of massive stars and cosmic rays, a high positron production rate $\approx 5 \times 10^{42}$ s$^{-1}$ is still required to explain the strong 511 keV line. Recently, there are some new scenarios proposed that can account for 100\% of the bulge 511 keV emission. For example, \citet{Crocker} propose that a single type of transient source, deriving from stellar populations of age 3-6 Gyr and yielding $\sim 0.03M_{\odot}$ of the positron emitter $^{44}$Ti, can simultaneously explain the strength and morphology of the 511 keV emission and the solar system abundance of the $^{44}$Ti decay product $^{44}$Ca. However, more observational evidence has to be obtained to support this claim. Therefore, we still need to explore other possibilities that can satisfactorily explain the 511 keV emission.

In the past decade, it was suggested that the positrons produced through dark matter annihilation can account for the 511 keV line and the high $B/D$ ratio, provided that the dark matter mass is of the order MeV \citep{Boehm,Ascasibar,Sizun}. Later studies show that the injection energy of the positrons should be as low as 3 MeV \citep{Beacom}. However, \citet{Wilkinson} find that the production of positrons by relic MeV dark matter annihilation violates the cosmological data. Therefore, this kind of proposal (MeV annihilating relic dark matter) is now disfavored. 

Recently, \citet{Chan} proposes a new annihilating dark matter model to account for the 511 keV line. By using the fact that there was a large dense cloud near the Milky Way center about $10^6$ years ago, he shows that the pair production inside the cloud due to high-energy photons injected by dark matter annihilation via $b\bar{b}$ channel can provide enough positrons to account for the strong 511 keV line. The mass of dark matter can be as large as $m=40$ GeV. This annihilating dark matter model is motivated by the dark matter interpretation of the GeV gamma-ray excess near the Milky Way center \citep{Daylan}. However, recent analyses based on the Fermi-LAT data of the Milky Way dwarf spheroidal satellite galaxies tend to disfavor this annihilation model (via $b\bar{b}$ channel) with $m \sim 40$ GeV \citep{Ackermann}. At the same time, some studies realize that the effect of inverse Compton scattering is significant near the Milky Way center \citep{Calore}. Some new possible annihilation channels (e.g. $\mu^+\mu^-$) have been suggested to account for the GeV excess. 

In this article, we follow the original idea of the pair-production mechanism in \citet{Chan} but extend our calculations to other different annihilation channels and dark matter mass. By using a standard dark matter density profile and considering the effect of the supermassive black hole, we perform a more detailed calculation and constrain the possible annihilation channels and dark matter mass, which can provide enough positrons to account for the strong 511 keV line and do not violate the current observational constraints.

\section{The pair-production model}
Recent studies show that a large amount of dense gas $\sim 10^5M_{\odot}$ in the form of a disk might exist near the Milky Way Center ($r \le 0.4$ pc) $10^{6.5}$ years ago. The dense cloud can provide enough gas to form the young and massive stars extending from 0.04 pc - 0.4 pc \citep{Wardle,Lucas,Wardle2}. It can also overcome tidal shear in the vicinity of the supermassive black hole and explain the truncation of the stellar surface density within 0.04 pc. The density and size of the dense cloud are $\sim 10^8$ cm$^{-3}$ and $5-7$ pc respectively \citep{Goicoechea,Zadeh}. Since most of the gas in the cloud was either captured by the supermassive black hole or converted to stars, this dense gas cloud cannot be found nowadays. 

In astrophysics, a large amount of positrons can be produced in the dense gas through pair-production mechanism ($\gamma \rightarrow e^++e^-$). If a photon has energy greater than $2m_ec^2$ and is entering the dense gas, a cascade of photons, electrons and positrons would be produced in the field of the nucleus from the surrounding gas. On the other hand, when a high-energy electron or positron is entering the dense gas cloud, it emits high-energy photons via Bremsstrahlung process. These high-energy photons can also generate a cascade of positrons via pair-production mechanism. The cross section for pair production is $\sigma_{pp} \approx 9 \times 10^{-27}$ cm$^2$ \citep{Longair}. 
 
Therefore, if dark matter annihilates and produce a large amount of high-energy photons and electron-positron pairs, a large amount of positrons can be produced via pair-production mechanism inside the dense cloud. Following \citet{Chan}, we assume that the average number density and the total size of the cloud are $n_g \sim 10^8$ cm$^{-3}$ and $R \sim 5$ pc respectively. The total optical depth of the electron-positron pair-production is $\tau \approx n_g \sigma_{pp} R \sim 13$ \citep{Chan}. This large optical depth can generate $\sim 100-1000$ positrons via pair-production mechanism for each high-energy photon, electron or positron \citep{Longair,Chan}. These positrons produced ($\sim 1$ MeV) would cool down to non-relativistic via synchrontron loss, inverse Compton scattering, bremsstrahlung loss and coulomb loss after leaving the dense cloud. The cooling time is of the order $10^6$ years \citep{Chan}. In other words, the positrons produced $\sim 10^6$ years ago by pair-production mechanism in the dense gas would use the same order of time ($10^6$ years) to cool down to non-relativistic and combine with hydrogen atoms to form positroniums, which consequently emit 511 keV photons. As a result, we can observe this 511 keV line nowadays even though the dense cloud disappears \citep{Chan}. 

The rate of dark matter annihilation within a radius $R$ is given by
\begin{equation}
\dot{N}_{DM}=\int_0^R \frac{\rho_{DM}^2}{m^2}<\sigma v>4 \pi r^2dr,
\end{equation}
where $\rho_{DM}$ is the density profile of dark matter. In \citet{Chan}, a generalized Navarro-Frenk-White (NFW) profile has been used to model the dark matter density profile. However, many studies indicate that the inner dark matter density profile would be steepened by the supermassive black hole near the Milky Way center so that a density spike would be resulted (the adiabatic growth model) \citep{Gondolo,Merritt,Fields}. The density spike can be modeled by the following form \citep{Fields}:
\begin{equation}
\rho_{DM}= \left \{ \begin{array}{lll}
0,       & {\ \ r \le 4GM_{BH}/c^2, } \\ &\\
\frac{\rho_{sp}(r)\rho_{in}(t,r)}{\rho_{sp}(r)+\rho_{in}(t,r)},       & {\ \ 4GM_{BH}/c^2 \le r \le r_b, } \\ &\\
\rho_b \left(\frac{r_b}{r} \right)^{\gamma_c},     & {\ \ r_b \le r \le r_s, } \end{array} \right.
\end{equation}
where $r_b=0.2GM/v_0^2$, $\rho_b=\rho_D(D/r_b)^{\gamma_c}$, $\rho_{sp}(r)=\rho_b(r_b/r)^{\gamma_{sp}}$, $\rho_{in}(t,r)=\rho_{ann}(t)(r/r_{in})^{-\gamma_{in}}$, $\gamma_{in}=1/2$ and $\gamma_{sp}=(9-2\gamma_c)/(4-\gamma_c)$. The density $\rho_{ann}=m/(<\sigma v>t)$ is called the annihilation plateau density and $r_{in}=3.1 \times 10^{-3}$ pc is the innermost radius of the spike. We take the following parameters for calculations: $v_0=105$ km/s, $D=8.5$ kpc, $\rho_D=0.008 M_{\odot}$ pc$^{-3}$, $M_{BH}=4 \times 10^6M_{\odot}$, $t=10^{10}$ yrs and $r_s=16$ kpc \citep{Fields}. By using the best-fit value reported in \citet{Calore,Daylan}, we take $\gamma_c=1.26$ (the best-fit value to account for the GeV excess). Here, we assume that the dark matter particles are relic so that we take the thermal relic annihilation cross section $<\sigma v>=2.2 \times 10^{-26}$ cm$^3$ s$^{-1}$ \citep{Steigman}. This value can obtain a correct cosmological dark matter abundance for thermal relic dark matter. Theoretically, dark matter can annihilate via different possible channels such as $e^+e^-$, $\mu^+\mu^-$ and $b\bar{b}$. Generally speaking, all of the annihilation channels can simultaneously produce photons, electrons and positrons. The total number of photons, electrons or positrons produced by dark matter annihilation is given by
\begin{equation}
\dot{N}=\int_0^m \dot{N}_{DM} \frac{dN'}{dE}dE,
\end{equation}
where $dN'/dE$ is the energy spectrum of the produced photons, electrons or positrons. The spectrums for different annihilation channels can be obtained in \citet{Cirelli}.

As mentioned above, each photon or electron-positron pair can produce a cascade of positrons via pair-production mechanism. This effect can be described by a boost factor $B(E,r)$. The boost factor depends on $r$ because the optical depth depends on the position of the photons or electron-positron pairs produced by dark matter annihilation. By using the pair-production model in \citet{Longair}, we rewrite the above equations and obtain the total number of positrons produced per second:
\begin{equation}
\dot{N}_{e^+}=\int_0^R \left[\frac{\rho_{DM}^2}{m^2}<\sigma v>4 \pi r^2 \int_0^m B(E,r) \frac{dN'}{dE}dE \right]dr.
\end{equation}
In Fig.~1, we show how $\dot{N}_{e^+}$ depends on $m$ for 12 popular annihilation channels ($e^+e^-$, $\mu^+\mu^-$, $\tau^+\tau^-$, $b\bar{b}$, $q\bar{q}$, $gg$, $\gamma \gamma$, $4e$, $4\mu$, $4\tau$, $W^+W^-$ and $ZZ$). Here, the symbol $q$ denotes a light quark ($u$, $d$ or $s$) and the $4e$ channel means dark matter annihilation first happens into some new boson $V$ which then decays into a pair of $e^+e^-$. As discussed in \citet{Prantzos}, the standard astrophysical sources such as supernovae, x-ray binaries and massive stars can explain about half of the strong 511 keV emission from the inner Milky Way. The remaining unexplained positron production rate is $\approx 5 \times 10^{42}$ s$^{-1}$. Based on the result in Fig.~1, we summarize the possible ranges of $m$ ($5-1000$ GeV) that can satisfy the remaining unexplained positron production rate in Table 1.

\begin{table}
\caption{Possible ranges of $m$ (within $5-1000$ GeV) that can produce $\dot{N}_{e^+}=5\times 10^{42}$ s$^{-1}$.}
 \label{table1}
 \begin{tabular}{@{}lc}
  \hline
  Annihilation channel &  $m$ (GeV) \\
  \hline
  $b\bar{b}$ & $\le 1000$ \\
  $e^+e^-$ & $\le 200$ \\
  $q\bar{q}$ & $\le 1000$ \\
  $gg$ & $\le 1000$ \\
  $\gamma \gamma$ & $\le 200$ \\
  $\mu^+\mu^-$ & $\le 100$ \\
  $\tau^+\tau^-$ & $\le 150$ \\
  $4e$ & $\le 200$ \\
  $4\mu$ & $\le 150$ \\
  $4\tau$ & $\le 200$ \\
  $W^+W^-$ & $\le 700$ \\
  $ZZ$ & $\le 700$ \\
  \hline
 \end{tabular}
\end{table}

\begin{figure*}
\vskip 5mm
 \includegraphics[width=120mm]{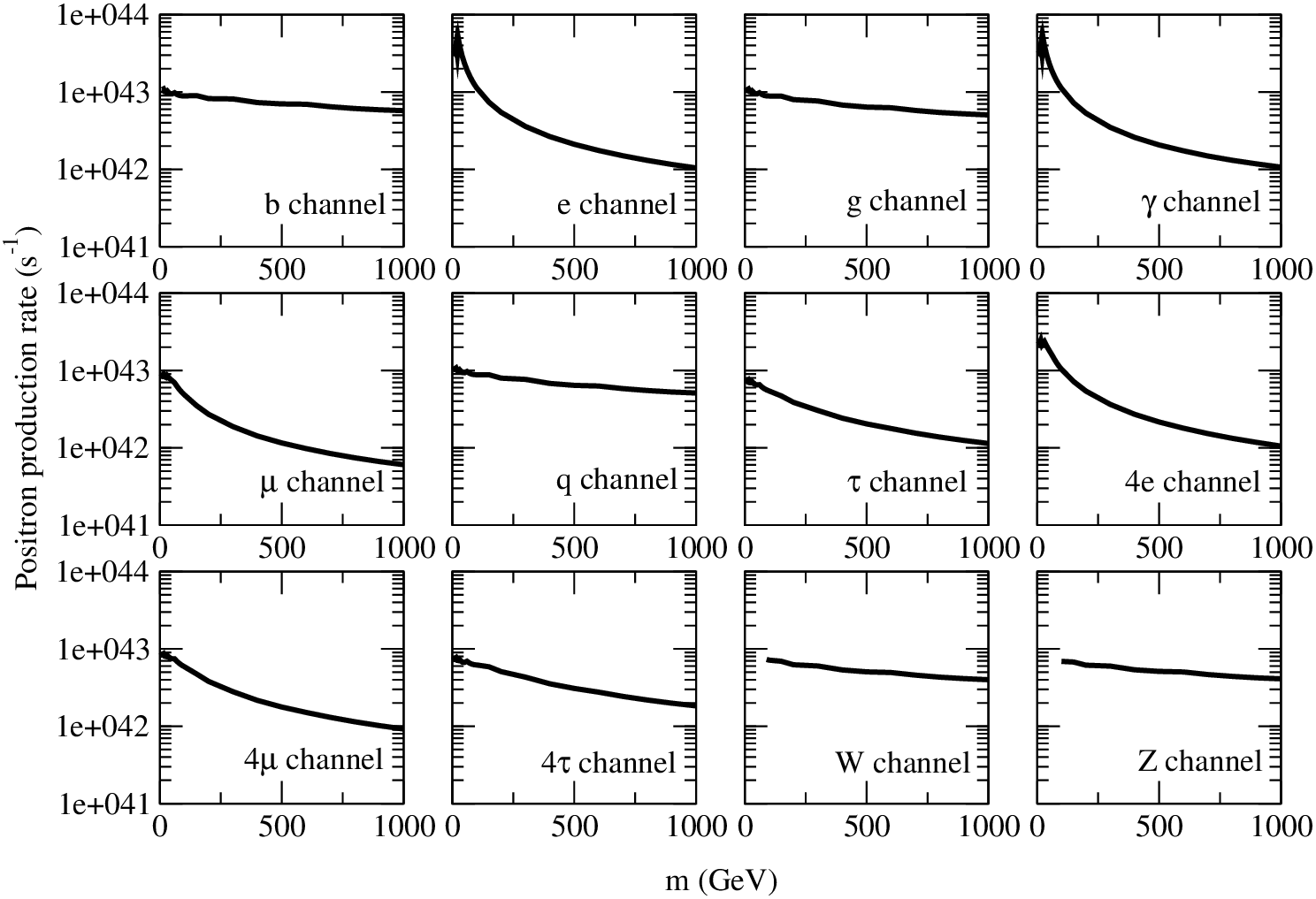}
 \caption{Positron production rate $\dot{N}_{e^+}$ versus $m$ for 12 popular annihilation channels. Here, we assume $<\sigma v>=2.2 \times 10^{-26}$ cm$^3$ s$^{-1}$.}
\vskip 5mm
\end{figure*}

Note that it is not easy for a positron to annihilate with free electrons inside the dense cloud to give photons. The probability of $e^+e^-$ annihilation inside the dense cloud is given by \citep{Prantzos}
\begin{equation}
P(E_i,E)=1-\exp \left[-n_e \int_{E}^{E_i} \frac{c \sigma_{ee}(E')dE'}{b(E')} \right],
\end{equation}
where $\sigma_{ee}(E') \sim 10^{-25}-10^{-30}$ cm$^2$ is the $e^+e^-$ annihilation cross section, $E_i$ is the injection energy of positrons before entering the dense cloud, $E$ is the final energy of positrons inside the dense cloud, $b(E')$ is the cooling rate and $n_e$ is the electron number density of the dense cloud. According to \citet{Wardle2}, the temperature of the dense cloud is below $T=2 \times 10^3$ K so that most of the particles are neutral hydrogen atoms or molecules. If we use a more conservative upper limit $T \le 5 \times 10^3$ K, by applying the Saha equation, we get $n_e \le 10^3$ cm$^{-3}$. Assuming $E_i=100$ GeV, the resulting probability $P(E_i,E)$ is less than 1\% for $E \ge 1$ MeV (see Fig.~2). This shows that the direct $e^+e^-$ annihilation inside the dense cloud is negligible. It is because the rate of energy loss $b(E')$ is very high inside the dense cloud (much higher than the annihilation rate). In Fig.~3, we also show the resultant energy spectrum of positrons for the $b\bar{b}$ channel just after leaving the dense cloud. We can see that most of the positrons have energy below 3 MeV, which satisfies the criterion suggested in \citet{Beacom}. Besides, the positrons produced would not annihilate with free electrons promptly after leaving the dense cloud. The cross section is too small for keV-MeV positrons to annihilate with free electrons ($\sigma_{ee} \sim 10^{-25}$ cm$^2$) due to the low density of interstellar medium ($n_e \sim 0.1-1$ cm$^{-3}$). The probability of $e^+e^-$ annihilation is less than 4\%. Therefore, most of the positrons produced would continually travel by $100-1000$ pc and further cool down to a very low energy. The cooling time scale is about $\sim 10^6$ years (cooling rate $\sim 10^{-13}-10^{-14}$ s$^{-1}$, see Fig.~4 for the cooled spectrum after $40000-80000$ years). Until the positron energy is below 100 eV, they would be much easier to form positroniums (cross section = $10^{-17}$ cm$^2$ for 100 eV positrons) and emit 511 keV photons consequently \citep{Chan}. Therefore, our model predicts that more than 95\% of the positrons (produced from the dense cloud) would become positroniums after travelling and cooling. This result is consistent with the fitted positronium fraction $f_P=96.7 \pm 2.2$\% \citep{Jean}.

\begin{figure*}
\vskip 5mm
 \includegraphics[width=120mm]{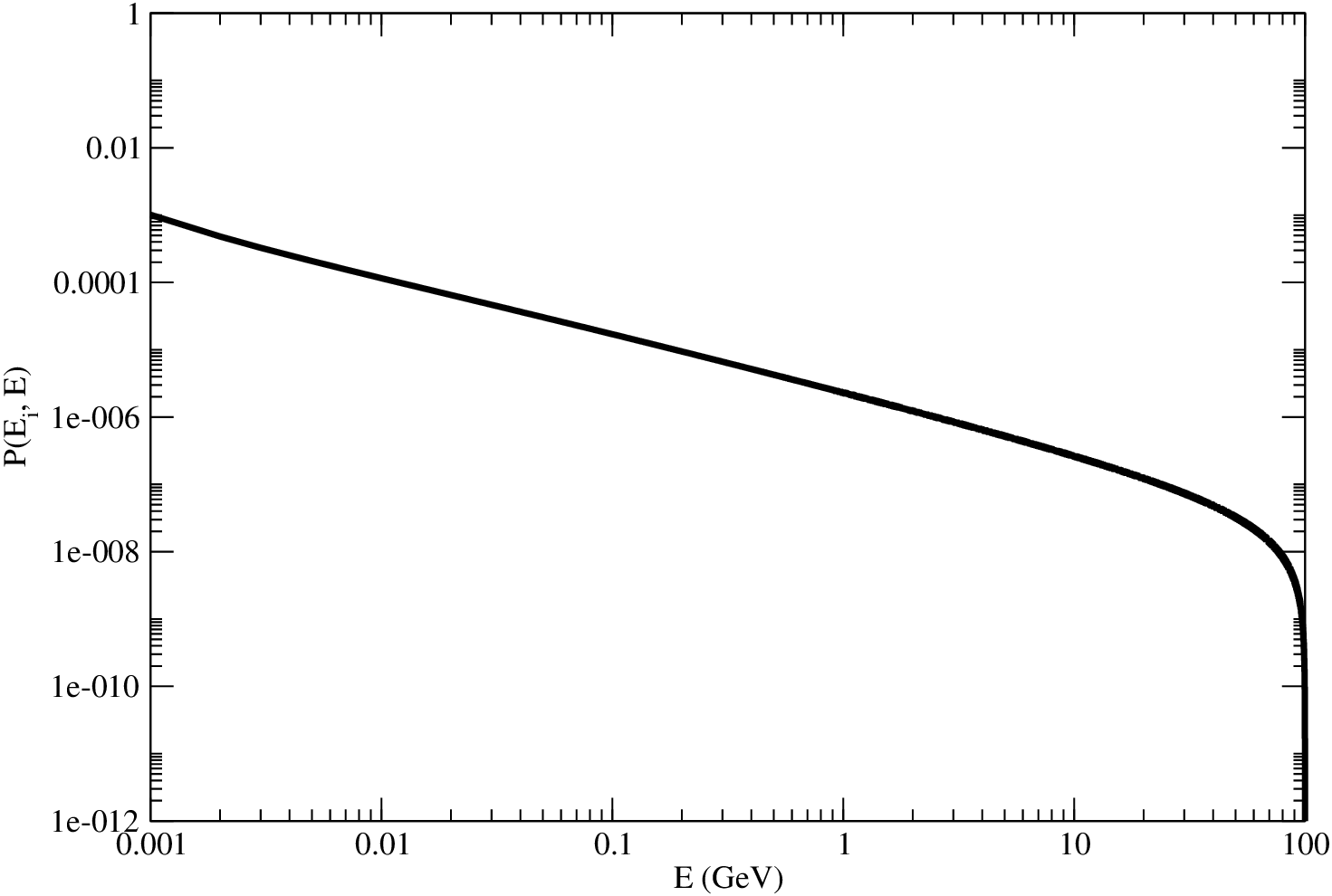}
 \caption{The probability $P(E_i,E)$ of $e^+e^-$ annihilation inside the dense cloud as a function of $E$. Here, we assume $E_i=100$ GeV.}
\vskip 5mm
\end{figure*}

\begin{figure*}
\vskip 5mm
 \includegraphics[width=120mm]{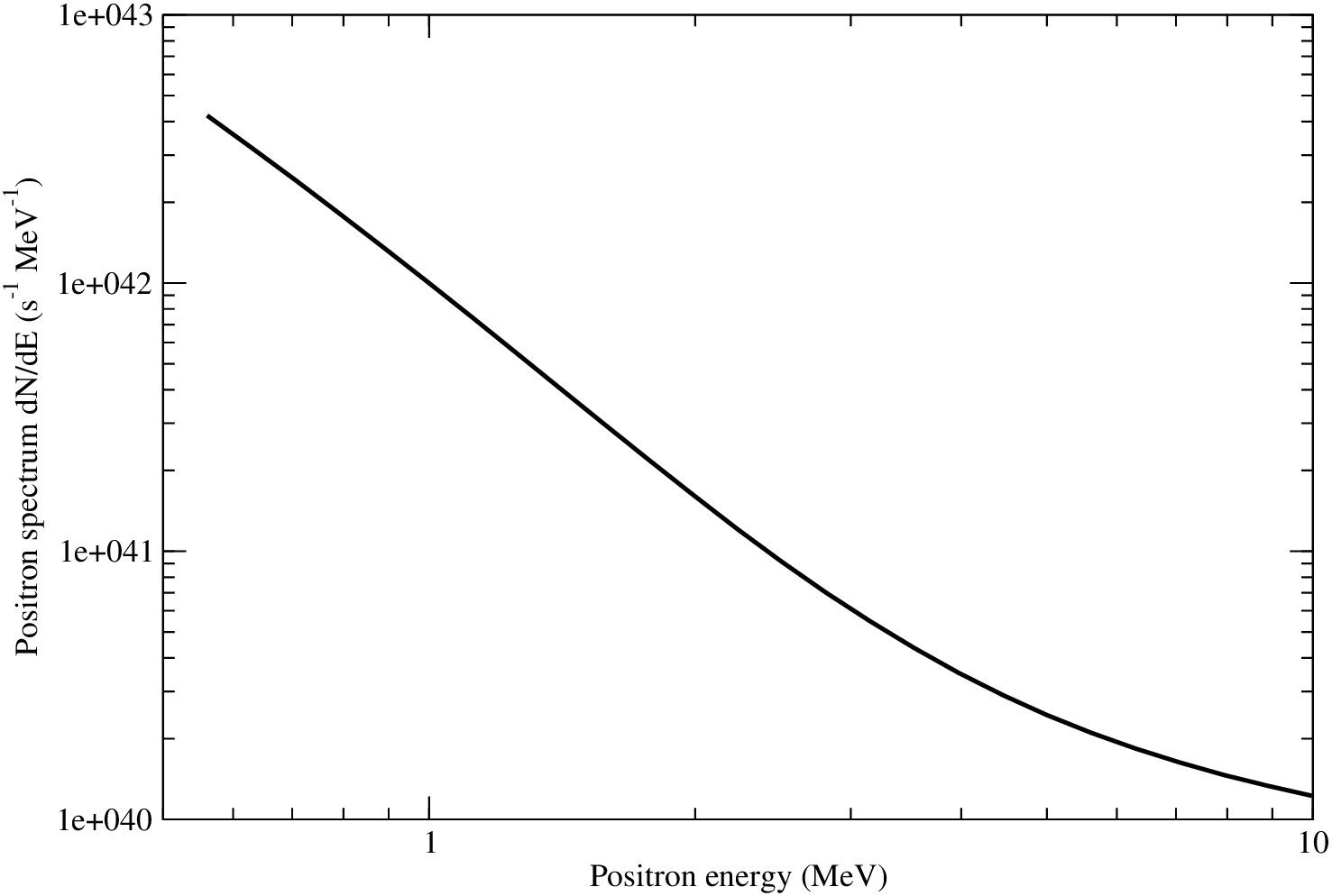}
 \caption{The resultant positron energy spectrum for the $b\bar{b}$ channel ($m=50$ GeV) after leaving the dense cloud. Here, we assume $<\sigma v>=2.2 \times 10^{-26}$ cm$^3$ s$^{-1}$.}
\vskip 5mm
\end{figure*}

\begin{figure*}
\vskip 5mm
 \includegraphics[width=120mm]{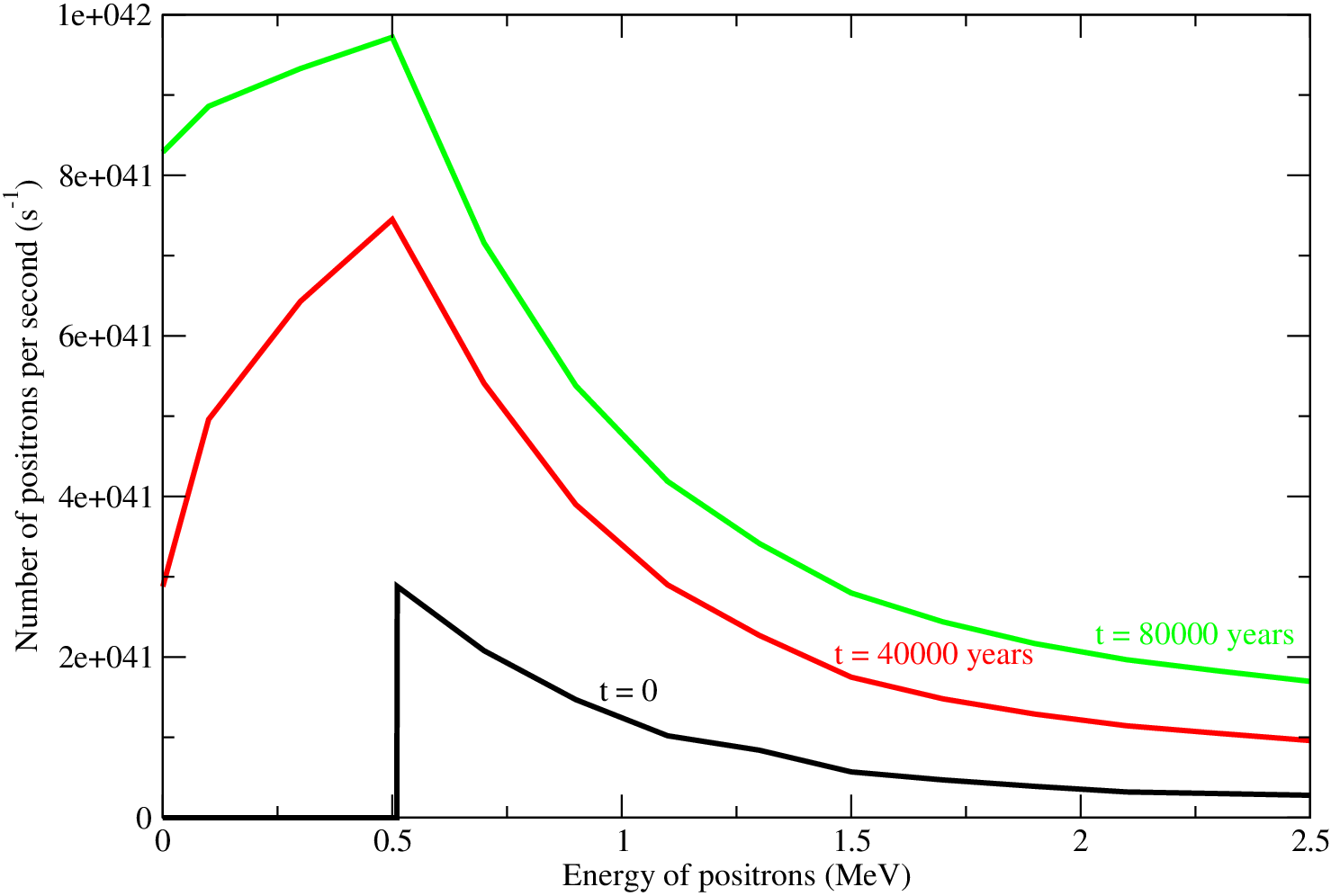}
 \caption{The cooled positron spectrum (averaged within the Galactic bulge) for the $b\bar{b}$ channel ($m=50$ GeV) after leaving the dense cloud ($t \le 80000$ years). Here, we assume that positrons are producing within the dense cloud during the time interval. We take $<\sigma v>=2.2 \times 10^{-26}$ cm$^3$ s$^{-1}$ and $n_e=1$ cm$^{-3}$.}
\vskip 5mm
\end{figure*}

\section{Other observational constraints and the combined results}
Besides the remaining unexplained positron production rate, our model should also satisfy the gamma-ray observational constraints. Since the dense cloud disappears, dark matter annihilation enhanced by the density spike would emit a strong gamma-ray flux which can be directly observed by us. The current observational constraint for gamma-ray flux ($1-100$ GeV) within $1^{\circ}$ of the Milky Way center is $3.2 \times 10^{-8}$ cm$^{-2}$ s$^{-1}$ \citep{Fields}. Recently, \citet{Fields} show that 40 GeV dark matter annihilating via $b\bar{b}$ channel gives a strong flux as large as $10^{-6}$ cm$^{-2}$ s$^{-1}$. Therefore, such an annihilation model has to be ruled out. 

We extend the calculations of the gamma-ray flux within $1^{\circ}$ ($R' \approx 150$ pc) for other annihilation channels. The gamma-ray flux due to dark matter annihilation is given by
\begin{equation}
\Phi=\frac{<\sigma v>}{8 \pi m^2}J \int \frac{dN_{\gamma}}{dE}dE,
\end{equation}
where $dN_{\gamma}/dE$ is the gamma-ray spectrum of dark matter annihilation and 
\begin{equation}
J= \int_{\Delta \Omega} \int_{\rm los} \rho_{DM}^2dl
\end{equation}
is the J-factor within a solid angle $\Delta \Omega$ along the line of sight $l$. Since we focus on a very small region within $1^{\circ}$, the J-factor can be approximately simplified to \citep{Ullio}
\begin{equation}
J \approx \frac{4\pi}{D^2} \int_0^{R'}r^2 \rho_{DM}^2dr.
\end{equation}
Therefore, the gamma-ray flux within $1^{\circ}$ can be given by $\Phi \approx \dot{N}/8 \pi D^2$. Note that this approximation scheme is not in general accurate. It is valid in our analysis because we are considering a very contracted profile $\gamma_{sp} \approx 2.36$. The error of the approximation would be a factor of 4 if an NFW profile is used. The calculated flux as a function of $m$ is shown in Fig.~5 for each channel. We also summarize the possible annihilation channels and mass ranges that can satisfy this stringent gamma-ray flux limit in Table 2.

By combining the results in Tables 1 and 2, we find that most of the channels are ruled out. Only 4 channels are able to account for the 511 keV line and satisfy the gamma-ray flux limit: $\mu^+\mu^-$ ($m=80-100$ GeV), $\gamma\gamma$ ($m=150-200$ GeV), $4e$ ($m=90-150$ GeV) and $4\mu$ ($m \le 150$ GeV). In particular, the range for the $\gamma \gamma$ channel is ruled out by the gamma-ray line detection \citep{Ackermann2}. Therefore, only three channels can satisfy the current constraints. Besides, we should pay more attention to the $\mu^+\mu^-$ annihilation channel. It is because recently \citet{Calore} show that the $\mu^+\mu^-$ channel with $m=60-70$ GeV (best-fit) can account for the Milky Way GeV excess if the effect of inverse Compton scattering is taken into account. Moreover, the best-fit range of the $\mu^+\mu^-$ channel is $m=88^{+31}_{-9}$ GeV for the dark matter interpretation of the AMS-02 data \citep{Mauro}. Surprisingly, these results are close to our range. In other words, dark matter annihilating via $\mu^+\mu^-$ channel with $m \sim 80$ GeV can simultaneously account for the 511 keV line, Milky Way positron excess and the GeV gamma-ray excess. This result is also compatible with the current gamma-ray observational constraint and the adiabatic growth model of supermassive black hole.

\begin{table}
\caption{Possible annihilation channels and mass ranges (within $5-1000$ GeV) that can satisfy the gamma-ray ($1-100$ GeV) flux limit within $1^{\circ}$ of the Milky Way center.}
 \label{table2}
 \begin{tabular}{@{}lc}
  \hline
  Annihilation channel &  $m$ (GeV) \\
  \hline
  $\gamma \gamma$ & $150-300$ \\
  $\mu^+\mu^-$ & $80-150$ \\
  $4e$ & $90-150$ \\
  $4\mu$ & $5-1000$ \\
  \hline
 \end{tabular}
\end{table}

\begin{figure*}
\vskip 5mm
 \includegraphics[width=120mm]{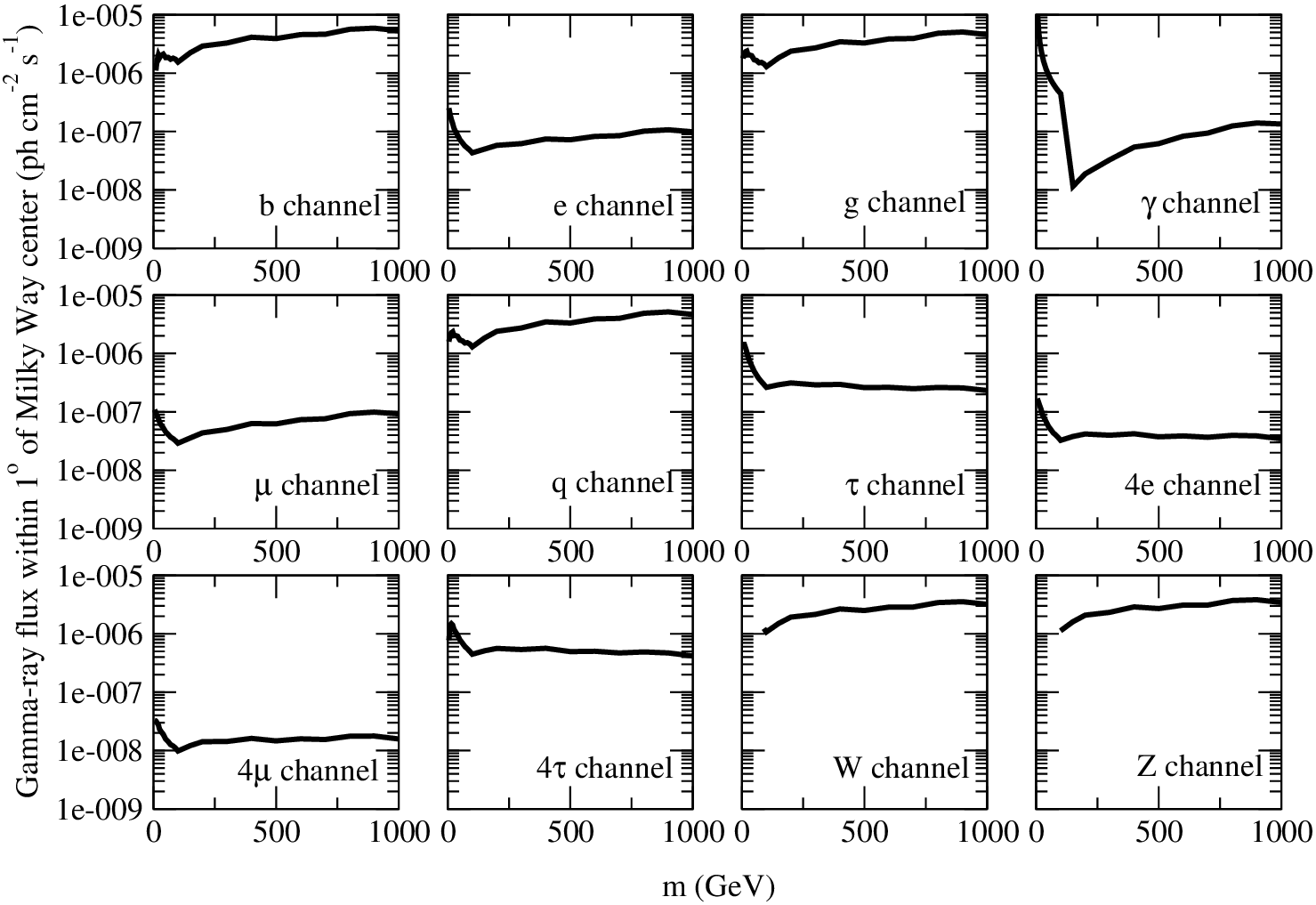}
 \caption{The gamma-ray flux (1-100 GeV) within $1^{\circ}$ of the Milky Way center versus $m$ for 12 popular annihilation channels. Here, we assume $<\sigma v>=2.2 \times 10^{-26}$ cm$^3$ s$^{-1}$.} 
\vskip 5mm
\end{figure*}

\section{Discussion and Conclusion}
In this article, we follow the pair-production model suggested in \citet{Chan} and perform a more detailed calculation to constrain the possible annihilation model and dark matter mass range. In this model, we first assume that a significant amount of high-energy positrons, electrons and photons are produced through dark matter annihilation. On the other hand, theoretical calculations show that a large dense cloud near the supermassive black hole used to exist $10^6$ years ago. These high-energy particles produced via dark matter annihilation inside this dense cloud would further produce a cascade of positrons by pair-production mechanism. The resultant rate of positron production in this model can be as high as $\dot{N}_{e^+} \sim 10^{43}$ s$^{-1}$, which can provide enough positrons to account for the 511 keV emission line. After leaving the dense cloud, these positrons produced would further cool down to $\sim 100$ eV by $\sim 10^6$ years. Only less than 4\% positrons would annihilate with free electrons during propagation. Therefore, more than 95\% cooled positrons would form positroniums with neutral hydrogen atoms and emit 511 keV photons, which agrees with observations. Also, since the size of the dense cloud ($\sim 5$ pc) is small relative to the Galactic bulge, it can be treated as a `point-source' production of positrons. These positrons would propagate outward and be deflected by the strong magnetic field near the Milky Way center. They would finally form positroniums randomly within the bulge. As a result, the 511 keV line emission would be close to spherically symmetric inside the bulge, which also agrees with observations. Therefore, GeV annihilating dark matter can provide enough positrons to explain the 511 keV line problem.

In our model, we also consider the effect of adiabatic contraction of dark matter density due to the supermassive black hole at the Milky Way center. Previous studies show that the adiabatic growth model of supermassive black hole is incompatible with the 40 GeV annihilating dark matter via $b\bar{b}$ channel \citep{Fields}. In fact, considering the adiabatic contraction of dark matter would affect the morphology within $\sim 2^{\circ}$ ($\approx 300$ pc) of the GeV excess (the contracted density profile would restore to the generalized NFW profile when $r \ge r_b=0.3$ pc). Many annihilation channels would have a significantly higher gamma-ray flux within $1^{\circ}$ of the Milky Way center. Therefore, the popular channels (e.g. $b \bar{b}$ and $\tau^+\tau^-$) which give a large amount of gamma rays would certainly fail under the assumption of the contracted dark matter density profile. However, the adiabatic growth model considered is the most natural one to describe the growth of supermassive black hole in our Milky Way \citep{Gondolo}. The quiet evolution for the disk in the Milky Way implies that the supermassive black hole formed is not due to merging process \citep{Wyse}. Therefore, although considering such a contracted density profile would give much more stringent limits for dark matter annihilation, the theoretical ground of this assumption is still strong. In our study, we find that only three annihilation channels ($\mu^+\mu^-$, $4e$ and $4 \mu$) with the thermal relic annihilation cross section can provide the required positron production rate and satisfy the stringent constraint of gamma-ray observations. In other words, we show that the dark matter annihilation scenario (via certain channels) is still compatible with the adiabatic growth model. In particular, the constrained mass range of the $\mu^+\mu^-$ channel is $m \sim 80$ GeV, which is close to the best-fit range that can account for the Milky Way GeV gamma-ray excess \citep{Calore}. Furthermore, this mass range agrees with the best-fit range ($m=88^{+31}_{-9}$ GeV) for the dark matter interpretation of the AMS-02 data \citep{Mauro}. It can also satisfy the Fermi-LAT constraint of the Milky Way dwarf spheroidal satellite galaxies ($m>6$ GeV for $\mu^+\mu^-$ channel) \citep{Ackermann}.

The results in this study are important and interesting. First of all, the $\mu^+\mu^-$ annihilation channel with dark matter mass $m \sim 80$ GeV and the thermal relic annihilation cross section can simultaneously account for the strong 511 keV line, GeV gamma-ray excess and the positron excess in the Milky Way. This model is also consistent with the adiabatic growth model of supermassive black hole. It can satisfy all of the current gamma-ray constraints, the observed positronium fractions and the required positron production rate to account for the 511 keV emission inside the bulge. Therefore, a single framework can bridge three different problems (the 511 keV line problem, the gamma-ray excess problem and the positron excess problem) together and satisfy all the required constraints. Future observations (e.g. DAMPE mission \citep{Gargano,Wang}) and direct detection experiments (e.g. PandaX-II \citep{Tan} and LUX \citep{Akerib} experiments) can further verify our model.

\section{acknowledgements}
We are grateful to the referee for helpful comments on the manuscript. This work is supported by a grant from The Education University of Hong Kong (Activity code: 04234).

\label{lastpage}

\end{document}